\documentclass[a4paper,11pt]{article}
\usepackage{pos}

\usepackage{physics}
\usepackage{amsmath}
\usepackage{bm} 
\usepackage{todonotes}
\usepackage{xspace}
\usepackage{float}
\usepackage{multicol} 
\usepackage{slashed}
\usepackage{tabularx}
\usepackage{cancel}
\usepackage{hyperref}

\usepackage{environ}

\usepackage{lineno} 

\def\inf {\ensuremath{\infty}\xspace}

\def\fm {\ensuremath{\text{fm}}\xspace}
\def\MeV {\ensuremath{\text{MeV}}\xspace}

\newcommand{\vect}[1]{\ensuremath{\bm{#1}}}

\NewEnviron{eq}{
    \begin{linenomath*}
        \begin{align}
            \BODY
        \end{align}
    \end{linenomath*}
}

\title{Progress on the exploratory calculation of the rare Hyperon decay $\Sigma^+ \to p \ell^+ \ell^-$}
\ShortTitle{$\Sigma^+ \to p \ell^+ \ell^-$ on the lattice}

\author*[a]{Felix Erben}
\author[a]{Vera G\"ulpers}
\author[a]{Maxwell T. Hansen}
\author*[a]{Raoul Hodgson}
\author[a]{Antonin Portelli}

\affiliation[a]{School of Physics and Astronomy, University of Edinburgh}

\emailAdd{raoul.hodgson@ed.ac.uk}
\emailAdd{felix.erben@ed.ac.uk}

\abstract{The rare Hyperon decay $\Sigma^+ \to p \ell^+ \ell^-$ is an $s \to d$ flavour changing neutral current process, which is highly suppressed within the Standard Model, and is therefore sensitive to new physics. Due to recent improvements in experimental measurements of this decay, the Standard Model theory prediction must also be improved in order to identify any new physics in this channel.
We present updates on our progress towards the first exploratory lattice calculation of the long-distance part of the form factors of this decay. This pilot calculation is performed on a 340 MeV pion mass ensemble using domain-wall fermions as part of the RBC-UKQCD collaboration.}

\FullConference{%
  The 39th International Symposium on Lattice Field Theory (Lattice2022),\\
  8-13 August, 2022 \\
  Bonn, Germany 
}

\begin{document}
\maketitle

\section{Introduction}
With the discovery of the Higgs boson completing the Standard Model (SM), the search for physics Beyond the Standard Model (BSM) has become a central focus of the particle physics community. Two key methods in this search are precision measurements of SM parameters and the search for enhancement of rare decays due to BSM physics.
The former attempts to find deviations of relations between parameters that are known to be exact within the SM, for example the relations implied by the unitarity of the CKM matrix. The latter works to identify new physics by measuring properties of decays that are rare or forbidden within the SM. Due to this suppressed SM background, new physics can have large relative effects. One example of this is the rare Hyperon decay, $\Sigma^+ \to p \ell^+ \ell^-$, which is the topic of these proceedings, and the similar rare Kaon decay $K^+ \to \pi^+ \ell^+ \ell^-$ \cite{Christ2015ProspectsDecays,Christ2015Longell,Christ2016Progressell-,Christ2016FirstKtopiell+ell-} which has seen recent progress with a physical point lattice QCD calculation \cite{Boyle:2022ccj}.

The rare Hyperon decay $\Sigma^+ \to p \ell^+ \ell^-$ is an $s \to d$ quark flavour changing neutral current process, meaning it can only occur at loop level within the SM and is therefore heavily suppressed. However, some BSM processes could allow for tree-level contributions to this decay, causing a potentially large relative enhancement. It is therefore crucial to have a SM prediction of this decay if it is to be used to search for such deviations. For full details of the theoretical work in these proceedings, see ref.~\cite{Erben:2022tdu}. 

\subsection{Experimental Motivation}
The rare Hyperon decay in the muonic channel $\Sigma^+ \to p \mu^+ \mu^-$ was first observed in 2005 by the HyperCP collaboration at Fermilab \cite{HyperCPCollaboration2005Evidencemu}. They isolated 3 events, allowing for a measurement of the branching fraction
\begin{eq}
    \label{eqn:HCP_measurement}
    \mathcal{B}(\Sigma^+ \to p \mu^+ \mu^-)_\text{HCP} = \left( {8.6^{+6.6}_{-5.4}}\pm 1.5\right) \times 10^{-8} \,,
\end{eq}
where the first error is statistical and the second systematic. In addition, the collaboration also reported a very tight clustering of these events in the dimuon invariant mass, being interpreted as possible evidence of a new intermediate particle of mass $214.3 \pm 0.5 \,\MeV$. This became known as the HyperCP anomaly.
The LHCb experiment made their first measurement of this decay in 2018 with $\mathcal{O}(10)$ events \cite{LHCb:2017rdd}, obtaining the branching fraction 
\begin{eq}
    \label{eqn:LHCb_measurement}
    \mathcal{B}(\Sigma^+ \to p \mu^+ \mu^-)_\text{LHCb} = \left(2.2^{+1.8}_{-1.3} \right) \times 10^{-8} \,.
\end{eq}
In addition, LHCb found no significant resonant structure in the dimuon invariant mass spectrum, providing no evidence for the HyperCP anomaly. Members of LHCb have expressed interest in improving this measurement with additional data and hopefully obtaining first measurements of angular observables, alongside potentially the first observation of the electronic decay mode $\Sigma^+ \to p e^+ e^-$ \cite{AlvesJunior2019ProspectsLHCb}.

\subsection{Theoretical Motivation}
\begin{figure}
    \centering
    \includegraphics{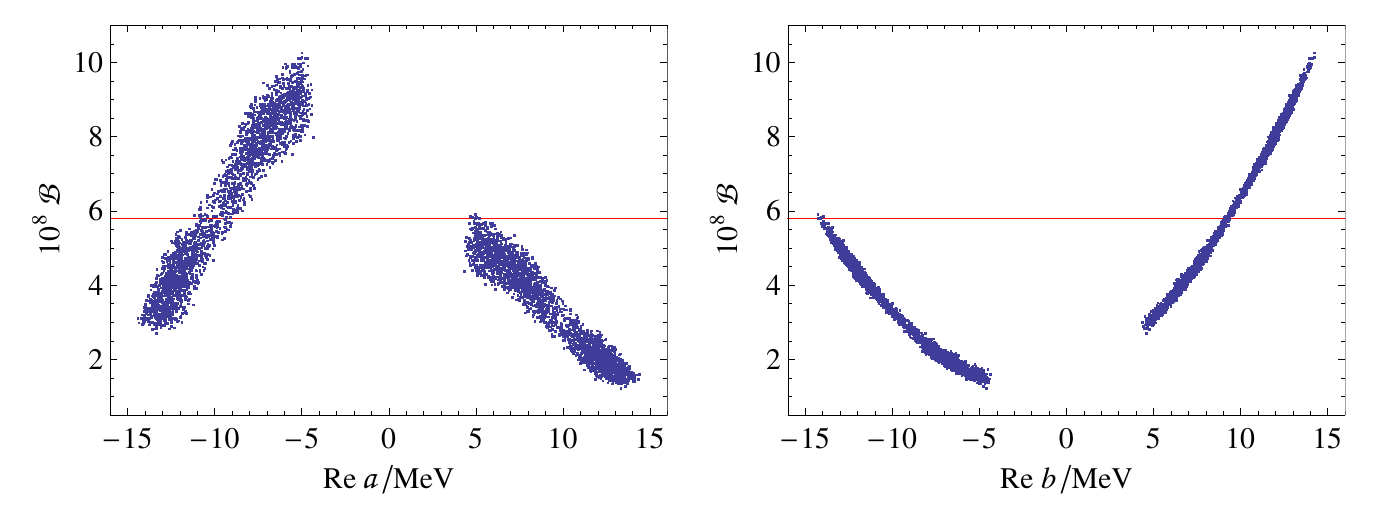}
    \caption{Branching fraction dependence of $\Sigma^+ \to p \mu^+ \mu^-$ on $\text{Re}(a)$ (left) and $\text{Re}(b)$ (right) from \cite{He2018DecayPmu+mu-}. The red line indicates the $2\sigma$ upper limit of the LHCb measurement in eq. \eqref{eqn:LHCb_measurement}.}
    \label{fig:He_ambiguity}
\end{figure}

The current SM prediction of the rare Hyperon decay shows that the short-distance contribution is $\sim 4$ orders of magnitude smaller than the long-distance contribution \cite{He2005DecayModel,He2018DecayPmu+mu-}, and therefore short-distance effects can be neglected at the current level of precision on the branching fraction. The dominant long-distance contribution comes from an intermediate virtual photon, producing the dilepton pair, $\Sigma^+ \to p \gamma^\ast \to p \ell^+ \ell^-$. Here the leptonic and hadronic parts of the amplitude factorise
\begin{eq}
    \mathcal{A}(\Sigma^+ \to p \ell^+ \ell^-) = -e^2 G_F \times \bar{u}_\ell(\vect{p}_{\ell^-}) \gamma^\mu v_\ell(\vect{p}_{\ell^+}) \times \mathcal{A}_\mu(\Sigma^+ \to p \gamma^\ast) \,,
\end{eq}
 leaving only 4 hadronic form factors to be determined,
\begin{eq}
    \mathcal{A}_\mu(\Sigma^+ \to p \gamma^\ast) = \bar{u}_p(\vect{p}) \left[ \left(a(q^2) + b(q^2) \gamma_5 \right) \sigma_{\mu \nu} q^\nu + \left(c(q^2)+d(q^2) \gamma_5 \right) \gamma_\mu \right] u_\Sigma(\vect{p}_\Sigma) \,.
\end{eq}
The current SM prediction of these form factors comes from a collection of methods \cite{He2005DecayModel,He2018DecayPmu+mu-,Geng:2021fog}. The real part of $a$ and $b$ can be constrained from experimental measurements of a similar decay with a real photon $\Sigma^+ \to p \gamma$. This does not, however, uniquely define $\text{Re}(a)$ and $\text{Re}(b)$, but instead it only does so up to a four-fold ambiguity. 
In addition, the real parts of the $c$ and $d$ form factors can be obtained from vector meson dominance models.
Finally, the imaginary parts of all 4 form factors are computed in Chiral Perturbation Theory ($\chi_\text{PT}$) using the optical theorem.
Putting these together gives the SM predictions \cite{He2005DecayModel,He2018DecayPmu+mu-,Geng:2021fog} of the branching fraction ($\mathcal{B}$) and forward-backward asymmetry ($A_\text{FB}$) of the $\Sigma^+ \to p \mu^+ \mu^-$ decay,
\begin{eq}
    1.6 \times 10^{-8} < \mathcal{B}(\Sigma^+ \to p \mu^+ \mu^-)_\text{SM} < 8.9 \times 10^{-8} \,, \\
    -1.4 \times 10^{-5} < A_\text{FB}(\Sigma^+ \to p \mu^+ \mu^-)_\text{SM} < 0.6 \times 10^{-5} \,,
\end{eq}
respectively. The large ranges in these values predominantly stem from the ambiguity in $\text{Re}(a)$ and $\text{Re}(b)$ along with differences from the possible choices of Baryon $\chi_\text{PT}$ that can be used. This can be seen in fig.~\ref{fig:He_ambiguity} where this ambiguity, in combination with the propagation of experimental uncertainty, gives relations between the unconstrained form factors and the branching fraction. For more details, see refs.~\cite{He2005DecayModel,He2018DecayPmu+mu-}.

It is clear from the large range in the SM prediction, that in order to constrain any new physics, our knowledge of these form factors must be improved. With the dominant source of uncertainty coming from the lack of constrain of $\text{Re}(a)$ and $\text{Re}(b)$ from the experimental input, any additional information about these form factors can be used to potentially improve the SM prediction. Even as little as an unambiguous determination of the sign of these form factors would allow for an improvement of the SM prediction (with one sign being more constraining than the other).

\section{Lattice Theory}
Lattice QCD provides us with a tool to make a systematically improvable ab initio determination of these form factors. This is done on a finite and discrete lattice in Euclidean space-time. For the purposes of these proceedings, we will ignore the issue of finite lattice spacing, and instead focus on the Euclidean space-time and finite volume (FV). For a detailed discussion of this work, see ref.~\cite{Erben:2022tdu}.

We start by defining the amplitude of interest in infinite-volume Minkowski space-time
\begin{eq}
    \mathcal{A}^{rs}_\mu(q) = \int d^4 x \bra{p(\vect{p}),r} T\left[ \mathcal{H}_W(x) J_\mu(0) \right] \ket{\Sigma^+(\vect{k}),s} \,,
\end{eq}
where $q = k-p$ is the 4-momentum transferred by the virtual photon, $J_\mu$ is the electromagnetic vector current, and $\mathcal{H}_W$ is the effective $s \to d$ weak Hamiltonian \cite{Buchalla1995WeakLogarithms}
\begin{eq}
    \mathcal{H}_W = \frac{G_F}{\sqrt{2}} V_{us} V_{ud}^\ast \left[ C_1 \left( Q_1^u - Q_1^c \right) + C_2 \left( Q_2^u - Q_2^c \right) + ... \right] \,,
    \label{eq:weak_Hamil}
\end{eq}
which is written in terms of the Wilson coefficients $C_i$, and 4-quark $V-A$ operators
\begin{eq}
    Q_1^q = [\bar{d} \gamma_\mu (1-\gamma_5) s] [\bar{q} \gamma^\mu (1-\gamma_5) q] \hspace{0.5cm} \text{and} \hspace{0.5cm}
    Q_2^q = [\bar{q} \gamma_\mu (1-\gamma_5) s] [\bar{d} \gamma^\mu (1-\gamma_5) q] \,,
\end{eq}
where the contraction within each bracket implies a sum over colors. There are additional operators that contribute to this weak Hamiltonian, however, their Wilson coefficients are suppressed by a factor $\left| \frac{V_{ts} V_{td}}{V_{us} V_{ud}} \right| \simeq 0.00142$, and we can therefore neglect these contributions.

Using the spectral representation, this amplitude can be written as
\begin{eq}
    \mathcal{A}^{rs}_\mu(q) & = i \int_0^\inf d\omega \left( \frac{\rho^{rs}_\mu(\omega)}{E_\Sigma(\vect{k}) - \omega + i \epsilon} - \frac{\sigma^{rs}_\mu(\omega)}{\omega - E_p(\vect{p}) - i \epsilon} \right) \,,
\end{eq}
in terms of two spectral functions $\rho$ and $\sigma$ which have contributions from states with strangeness quantum number $0$ and $-1$ respectively, and $E_B(\vect{p})$ is the energy of the baryon $B$ with 3-momentum $\vect{p}$. For convenience on the lattice, we also define a set of Dirac matrix valued quantities denoted with a tilde, for example,
\begin{eq}
    \mathcal{A}^{rs}_\mu(q) = \bar{u}_p^r(\vect{p}) \widetilde{\mathcal{A}}_\mu (q) u_\Sigma^s(\vect{k}) \,.
\end{eq}

In the finite-volume (FV) Euclidean space-time, we must first consider a correlation function similar to the form of our amplitude,
\begin{eq}
    \Gamma^{(4)}_\mu(t_p,t_H,t_\Sigma) & = \int d^3 \vect{x} \, \langle \psi_p(t,\vect{p}) \, \mathcal{H}_W(t_H,\vect{x}) J_\mu(0) \, \bar{\psi}_\Sigma(t_\Sigma,\vect{k}) \rangle \,,
\end{eq}
where $\psi_\Sigma(t,\vect{p})$ and $\psi_p(t,\vect{p})$ are operators that have the quantum numbers of a $\Sigma^+$ and proton respectively with momentum $\vect{p}$. We define the amputated 4-point function by removing the overlaps and propagation of the external states
\begin{eq}
    \hat{\Gamma}^{(4)}_\mu(t_H) & = \Gamma^{(4)}_\mu(t_p,t_H,t_\Sigma) \frac{E_p(\vect{p}) E_\Sigma(\vect{k})}{Z_p(\vect{p}) Z^\ast_\Sigma(\vect{k}) M_p M_\Sigma} e^{+E_p(\vect{p}) t_p} e^{-E_\Sigma(\vect{k}) t_\Sigma} \,,
    \label{eq:amputated_4point}
\end{eq}
where the interpolator overlap is given by $Z_B(\vect{p}) u_B^s(\vect{p}) = \bra{0} \psi_B(0) \ket{B(\vect{p}),s}$, and $M_B$ is the mass of baryon $B$. These quantities can all be extracted from the relevant 2-point correlation functions. Ground state dominance of the initial and final states has been assumed, leaving $\hat{\Gamma}^{(4)}(t_H)$ as only a function of the weak Hamiltonian time $t_H$, and not $t_\Sigma$ or $t_p$. 
The amputated 4-point function has the spectral representation
\begin{eq}
    \hat{\Gamma}^{(4)}_\mu(t_H) & = \int_0^\inf d\omega \, \mathbb{P}_p(\vect{p}) \, \left\{ \begin{array}{lcl}
        \widetilde{\sigma}(\omega)_L \, e^{-(\omega-E_p(\vect{p})) t_H} & \mbox{for}
        & t_H > 0 \\
        \widetilde{\rho}(\omega)_L \, e^{-(E_\Sigma(\vect{k})-\omega) t_H} & \mbox{for} & t_H < 0
        \end{array}\right\} \, \mathbb{P}_\Sigma(\vect{k}) \,,
\end{eq}
where we have introduced the projectors $\mathbb{P}_B(\vect{p}) = \frac{1}{2M_B} \sum_s u^s_B(\vect{p}) \bar{u}^s_B(\vect{p}) = \frac{-i \slashed{p}_B + M_B}{2M_B}$. 
The FV spectral densities $\rho_\mu(\omega)_L$ and $\sigma_\mu(\omega)_L$ are described by the matrix elements,
\begin{eq}
    \rho^{rs}(\omega)_L & = \sum_n \frac{\delta(\omega-E_n(\vect{k}))}{2 E_n(\vect{k})} \bra{p(\vect{p}),r} J_\mu(0) \ket{E_n(\vect{k})}_L \bra{E_n(\vect{k})} \mathcal{H}_W(0) \ket{\Sigma(\vect{k}),s}_L \,, \\
    \sigma^{rs}(\omega)_L & = \sum_m \frac{\delta(\omega-E_m(\vect{p}))}{2 E_m(\vect{p})} \bra{p(\vect{p}),r} \mathcal{H}_W(0) \ket{E_m(\vect{p})}_L \bra{E_m(\vect{p})} J_\mu(0) \ket{\Sigma(\vect{k}),s}_L \,,
\end{eq}
where the sum is over the discrete set of FV states.
Integrating this amputated 4-point function within the window $t_H \in [-T_a,T_b]$ gives the object
\begin{eq}
    I^{(4)}_\mu(T_a,T_b) & = -i \int_{-T_a}^{T_b} dt_H \, \hat{\Gamma}^{(4)}_\mu (t_H) \\
    & = i \int_0^\inf d\omega \, \left[ \widetilde{\rho}_\mu(\omega)_L \frac{1-e^{(E_\Sigma(\vect{k})-\omega)T_a}}{E_\Sigma(\vect{k})-\omega} - \widetilde{\sigma}_\mu(\omega)_L \frac{1-e^{-(\omega-E_p(\vect{p}))T_b}}{\omega-E_p(\vect{p})} \right] \,.
\end{eq}
It can be seen that this function has similar structure to the amplitude of interest (with the spectral functions replaced with their FV counterpart), and some exponential pieces depending on $T_a$ and $T_b$. Clearly, for any regions where $\omega > E_\Sigma(\vect{k})$ and $\omega > E_p(\vect{p})$, this object has a well-defined $T_a,T_b \to \inf$ limit where the exponentials simply vanish. However, for any regions where $\omega < E_\Sigma(\vect{k})$, the first term will diverge as $T_a \to \inf$, and similar for the second term if $\omega < E_p(\vect{p})$.

By examining the spectrum of the $\sigma$ function, we can see that the intermediate states must have non-zero strangeness and momentum $\vect{p}$, and therefore $\omega > E_p(\vect{p})$ over the whole spectrum. Therefore, the $T_b \to \inf$ limit can be taken without complication. The $\rho$ spectrum, however, does contain states with $\omega < E_\Sigma(\vect{k})$. At the physical point these are the single proton state $\ket{p(\vect{k})}$, and the nucleon-pion like multiparticle FV states with total momentum $\vect{k}$. In order to take the $T_a \to \inf$ limit, we must first remove these growing exponentials. This can be done by noting that the energies and matrix elements involved in these problematic terms can be extracted from appropriate 2- and 3-point functions. In addition, in practice we cannot take $T_a,T_b \to \inf$ on a finite lattice, and therefore it may also be beneficial to remove a set of the decaying exponentials as well to improve the convergence to these limits. We define the FV estimator as the integrated amputated 4-point function with all exponential terms removed
\begin{eq}
    \widetilde{F}_\mu(\vect{k},\vect{p})_L = & \lim_{T_a,T_b \to \inf} \left[ I_\mu^{(4)}(T_a,T_b) \right. \\
    & \left. + i \sum_{n=0}^N \frac{\bra{p(\vect{p}),r} J_\mu(0) \ket{E_n(\vect{k})}_L \bra{E_n(\vect{k})} \mathcal{H}_W(0) \ket{\Sigma(\vect{k}),s}_L}{2 E_n(\vect{k})} \frac{e^{(E_\Sigma(\vect{k})-E_n(\vect{k}))T_a}}{E_\Sigma(\vect{k})-E_n(\vect{k})} \right. \nonumber \\
    & \left. -i \sum_{m=0}^M \frac{\bra{p(\vect{p}),r} \mathcal{H}_W(0) \ket{E_m(\vect{p})}_L \bra{E_m(\vect{p})} J_\mu(0) \ket{\Sigma(\vect{k}),s}_L}{2 E_m(\vect{p})} \frac{e^{-(E_m(\vect{p})-E_p(\vect{p}))T_b}}{E_m(\vect{p})-E_p(\vect{p})} \right] \,. \nonumber
\end{eq}
where the upper limit $N$ must be such that $E_N(\vect{k}) \geq E_\Sigma(\vect{k})$.
There is an equivalent procedure where the states are first removed from the unintegrated 4-point function $\Gamma_\mu^{(4)}$, leaving the integration over all time convergent without the need for regularising with $T_a,T_b$. This removal causes these states to not be present after the integral, so their contribution to the FV estimator must be added back using the objects from the 2- and 3-point functions again. These two methods are algebraically identical, but may differ numerically when performing an analysis at finite statistics. For details of this alternative approach, see ref.~\cite{Erben:2022tdu}.

Finally, for the single-proton intermediate state, there exists another method that will remove this contribution from the 4-point function, without affecting the FV estimator. This is done by adding additional operators to the weak Hamiltonian which cancels the contribution from this state, but affects the matrix elements of the rest of the spectrum to compensate for this missing state.

Consider the $s \to d$ flavour changing scalar and pseudoscalar currents $\mathcal{S}=\bar{d}s$ and, $\mathcal{P}=\bar{d} \gamma_5 s$ respectively. It can be seen that, due to the chiral Ward identities, the FV estimator is unaffected by the exchange $\mathcal{H}_W \to \mathcal{H}_W' = \mathcal{H}_W + c_S \mathcal{S} + c_P \mathcal{P}$ in the 4-point function. However, tuning the arbitrary coefficients $c_S$ and $c_P$ such that the matrix element $\bra{p(\vect{k})} \mathcal{H}_W' \ket{\Sigma(\vect{k})} = 0$ removes the growing exponential term coming from the single proton intermediate state in $I_\mu^{(4)}(T_a,T_b)$. This process has been discussed in the case of the rare Kaon decay $K \to \pi \ell^+ \ell^-$ previously in  refs.~\cite{Christ2015ProspectsDecays,Christ2015Longell,Christ2016Progressell-,Christ2016FirstKtopiell+ell-}, but due to the additional spin degree of freedom of the baryons, the weak Hamiltonian matrix element has two form factors (unlike the rare Kaon decay that has only one)
\begin{eq}
    \bra{p(\vect{k}),r} \mathcal{H}_W(0) \ket{\Sigma^+(\vect{k}),s} = \bar{u}_p^r(\vect{k}) \left[ a_H + b_H \gamma_5 \right] u_\Sigma^s(\vect{k}) \,,
\end{eq}
requiring both the scalar and pseudoscalar shifts to remove this state in general.
There is, however, a special kinematic point of the $\Sigma^+$ at rest, $\vect{k} = \vect{0}$, where the spinor contraction $\bar{u}_p^r(\vect{0}) \gamma_5 u_\Sigma^s(\vect{0}) = 0$, and therefore the pseudoscalar shift is no longer required.

From these methods above, in theory all exponentially growing contributions (and as many higher states as desired) can be removed, giving a measurement of the finite-volume estimator $\widetilde{F}_\mu(\vect{k},\vect{p})_L$.

\section{Finite Volume Effects}

The discrete spectrum of finite-volume $N \pi$ states leads to poles in \begin{align}
F^{rs}_\mu(\boldsymbol{k},\boldsymbol{p})_{L} = \bar{u}^r_p(\boldsymbol{p}) \widetilde{F}_\mu(\boldsymbol{k},\boldsymbol{p})_{L} u^s_\Sigma(\boldsymbol{k}) \, .
\end{align}
Recovering the physical amplitude 
\begin{align}
\mathcal{A}^{rs}_\mu(k,p) & = F^{rs}_\mu(\boldsymbol{k},\boldsymbol{p})_{L} + \Delta F^{rs}_{\mu}(\boldsymbol{k},\boldsymbol{p})_L \,,
\end{align}
necessitates estimating the finite-volume correction term $\Delta F^{rs}_{\mu}(\boldsymbol{k},\boldsymbol{p})_L $, containing the exact same poles as $ F^{rs}_\mu(\boldsymbol{k},\boldsymbol{p})_{L} $ so that they cancel and $\mathcal{A}^{rs}_\mu(k,p)$ is pole-free.
This correction term can be written \cite{Kim:2005gf,Christ:2015aha,Briceno:2019opb}
\begin{align}
\label{eq:DeltaFDef}
\Delta F^{rs}_\mu(\boldsymbol{k},\boldsymbol{p})_L & = i \mathcal A^r_{J_\mu}(E_{\Sigma}(\boldsymbol k), \boldsymbol{k},\boldsymbol{p}) \cdot \mathcal{F}\big ( E_{\Sigma}(\boldsymbol k), \boldsymbol{k},L \big ) \cdot \mathcal A^s_{H_W}(E_{\Sigma}(\boldsymbol k), \boldsymbol k) \,,
\end{align}
with the three transition amplitudes
\begin{align}
\mathcal A^r_{J_\mu}(E_{\Sigma}(\boldsymbol k), \boldsymbol{k},\boldsymbol{p}) &=  \big<N(\boldsymbol{p}),r\big|J_\mu(0)\big|E, (N \pi)^\mathrm{in}(\boldsymbol{k} )\big> \,, \\
\mathcal A^s_{H_W}(E_{\Sigma}(\boldsymbol k), \boldsymbol k) &=   \big<E, (N \pi)^\mathrm{out}(\boldsymbol{k} )\big| \mathcal H_W(0) \big|{\Sigma}(\boldsymbol{k}),s\big>  \,, \\
\mathcal{F}(E, \boldsymbol{P},L) & = \frac{1}{F(E, \boldsymbol{P},L)^{-1}+\mathcal{M}(E_{\sf cm})} \,.
\label{eq:calFdef}
\end{align}
For two non-identical scalar particles with masses $M_\pi, M_N$, the geometric function $F(E, \boldsymbol{P},L)$ can be written as a sum-integral difference,
\begin{align}
F_{\ell' m';\ell m}(E, \boldsymbol P,L)= \bigg[ \frac{1}{L^3} \sum_{\boldsymbol{k}} - \int \frac{\dd^3\boldsymbol{k}}{(2\pi)^3} \bigg] f_{\ell' m';\ell m}(E, \boldsymbol P, \boldsymbol k, M_\pi, M_N) \,,
\end{align}
with a known function $f_{\ell' m';\ell m}(E, \boldsymbol P, \boldsymbol k, M_\pi, M_N)$ that is given in \cite{Erben:2022tdu}. A projection with suitable Clebsch-Gordon coefficients promotes this equation to the case of particles with spin~\cite{Briceno:2014oea}, 
\begin{equation}
F_{J' \ell' \mu', J \ell \mu}(E, \boldsymbol P, L) = \sum_{m,\sigma, m'} \langle \ell m;\frac 12 \sigma | J \mu \rangle \langle \ell' m';\frac 12 \sigma | J' \mu' \rangle F_{\ell'm' ;\ell m}(E, \boldsymbol P, L) \,,
\end{equation}
with quantum numbers for total angular momentum $J$ and its azimuthal component $\mu$, orbital angular momentum, $\ell$, and total spin $s=1/2$. Using those same quantum numbers, the $N \pi$ scattering amplitude can be written
\begin{align}
\mathcal{M}_{J' \ell '\mu';J \ell \mu}(E_{\sf cm}) =\delta_{J' J} \delta_{l' l}\delta_{\mu' \mu} \frac{8 \pi E_{\sf cm}}{p \cot \delta_{J,\ell}(p) - ip} \,,
\end{align}
with the centre-of-mass energy $E_{\sf cm}^2 = E^2 - \boldsymbol P^2$ and the scattering phase shift $\delta_{J,\ell}(p)$.

Near the finite-volume energies $E_n$, the Lellouch-Lüscher formalism \cite{Lellouch:2000pv} and its extensions lead to
\begin{equation}
\lim_{E \to E_n(L)} \big (E - E_n(L) \big )\frac{1}{ \mathcal{M}(E^{\sf cm}_n) + F^{-1}(E_n, \boldsymbol P,L) } = \mathcal E^{(n), {\sf in}} \otimes \mathcal E^{(n), {\sf out}} \,,
\end{equation}
where the $\otimes$ indicates an outer-product in the $J, \ell, \mu$ index space. This determines the transition amplitudes
\begin{align}
  \big < E_{n}, \boldsymbol k \big | {\mathcal H_{W}}(0) \big|{\Sigma}(\boldsymbol{k}),s\big>_L & = \mathcal E^{(n), {\sf out}} \cdot \mathcal A^s_{H_W}(E, \boldsymbol k)  \,, \\
 \big<N(\boldsymbol{p}),r\big|J_\mu(0)\big|E_{n}, \boldsymbol k \big>_L & =  \mathcal A^r_{J_\mu}(E, \boldsymbol k,\boldsymbol p) \cdot \mathcal E^{(n), {\sf in}} \,.
\end{align}
Given full knowledge of the $N \pi$ finite-volume energy spectrum and the matrix elements in the above equations, the finite-volume correction $\Delta F^{rs}_{\mu}(\boldsymbol{k},\boldsymbol{p})_L $ can be determined. In the simplest case of a single channel and total momentum $\boldsymbol P = \boldsymbol 0$, the equations can be parameterised by the scattering phase shift $\delta^{\ell=1}(E)$ and the so-called pseudophase $\phi(E,L)$ via
\begin{equation}
\mathcal{M}(E) = \frac{8 \pi E }{p} \frac{1}{\cot \delta(E ) -i} \, , \qquad
F(E,L) = \frac{p}{8 \pi E } \, \big [ \! \cot \phi(E,L) + i \big ]
\, ,
\end{equation}
leading to
\begin{multline}
  \Delta F^{rs}_\mu(\boldsymbol{0},\boldsymbol{p})_L =  - i \frac{p_{\Sigma}}{8 \pi M_{\Sigma} } \mathcal A^r_{J_\mu}(M_{\Sigma}, \boldsymbol 0,\boldsymbol p) e^{- 2 i \delta(M_{\Sigma})} \mathcal A^s_{H_W}(M_{\Sigma}, \boldsymbol 0) \Big ( \! \cot \!\big [ \delta(M_{\Sigma} )+ \phi(M_{\Sigma},L) \big ]+i \Big ) \,,
  \label{eq:dFgi_final}
  \end{multline}
  matching eq.~(35) of ref.~\cite{Christ:2015pwa}. This result is vastly simpler than the result for $\boldsymbol P \neq \boldsymbol 0$ and we therefore plan to limit our first calculations to the case of the $\Sigma^+$ at rest. This is not particularly limiting, as a momentum injection into the $p$ leads to a larger kinematic range at small lattice momenta, which is preferable and allows access to an interesting range of the spectrum. 
  
  In ref.~\cite{Erben:2022tdu} we show that indeed this quantity leads to an amplitude free from singularities, i.e. that for an expansion around $\delta L = \bar{L}-L$, where $\bar{L}$ is some tuned volume such that one finite-volume energy obeys $E_{\bar{n}}=M_\Sigma$, we get
  \begin{align} F^{rs}_\mu(\boldsymbol{0},\boldsymbol{p})_L = \mathcal{O}(\delta L^{-1}), \ \ 
    \Delta F^{rs}_\mu(\boldsymbol{0},\boldsymbol{p})_L  = \mathcal{O}(\delta L^{-1}), \ \ 
    F^{rs}_\mu(\boldsymbol{0},\boldsymbol{p})_L+
    \Delta F^{rs}_\mu(\boldsymbol{0},\boldsymbol{p})_L  = \mathcal{O}(\delta L^0) \, .
  \end{align}
We also give the first $\mathcal{O}(\delta L^1)$ correction term to this result in ref.~\cite{Erben:2022tdu}.

\section{Numerical Study}
With the formalism for extracting the rare Hyperon decay amplitude from the lattice known, a first exploratory numerical calculation is currently in progress. For this study, we use a 2+1 flavour domain-wall-fermion ensemble from the RBC-UKQCD collaboration with a lattice spacing of $a \simeq 0.1 \,\fm$, a spatial extent of $L \simeq 2.7 \,\fm$, and a Pion mass of $M_\pi = 340 \,\MeV$ \cite{RBC:2010qam}. In addition, the baryons masses have been measured to be $M_N \simeq 1150 \,\MeV$ and $M_\Sigma \simeq 1350 \,\MeV$ for the nucleon and $\Sigma$ respectively.
This lattice was chosen, in part, to reduce the energy gap between the Pion and baryons, which reduces the exponential signal-to-noise problem of the baryonic correlation functions. In addition, it can be seen that the $N \pi$ threshold is above the $\Sigma$ energy and therefore the $N \pi$-like FV intermediate states will decay exponentially in $T_a$. This leaves only the single proton intermediate state to be removed before taking $T_a \to \inf$. In addition, there will be no powerlike FV corrections on this ensemble because these stem from the multiparticle FV states below the $E_\Sigma(\vect{k})$ threshold, of which there are none with these masses. Collectively, these features make this ensemble excellent for a proof-of-principal calculation of this decay.

For the Wilson coefficients in eq.~\eqref{eq:weak_Hamil}, we use those calculated in ref.~\cite{Christ:2012se} on a similar ensemble with the same lattice spacing and action, but smaller volume and larger Pion mass. We expect that the Wilson coefficients will be unaffected to the level of precision required here, since they are ultraviolet quantities and shouldn't be strongly affected by infrared physics.

To perform the measurements of the relevant correlation functions, we use the Grid \cite{Boyle2016Grid} and Hadrons \cite{Portelli2020Hadrons} C++ libraries, where all the necessary contraction code has been implemented. We use gauge-fixed Gaussian sources for the baryonic interpolators to improve the overlap with the ground state. We focus on the kinematic point with the $\Sigma^+$ at rest ($\vect{k} = \vect{0}$) and the proton with a single unit of lattice momentum $\vect{p}=\frac{2\pi}{L}(1,0,0)$. With this setup, we are able to utilise the scalar shift method of removing single proton state without the need for the pseudoscalar shift. 

We take measurements on 70 configurations, and in order to gain extra statistics, we measure on 32 time translation of each configuration. In order to improve our approximate momentum projection, we use a method known as field sparsening \cite{Li:2020hbj}, which we refer to in this context as source-sink sampling, and is described in ref.~\cite{Hodgson:2021lzt}. We use 4 samples per configuration at both the source and sink in this work.

\begin{figure}
    \centering
    \includegraphics{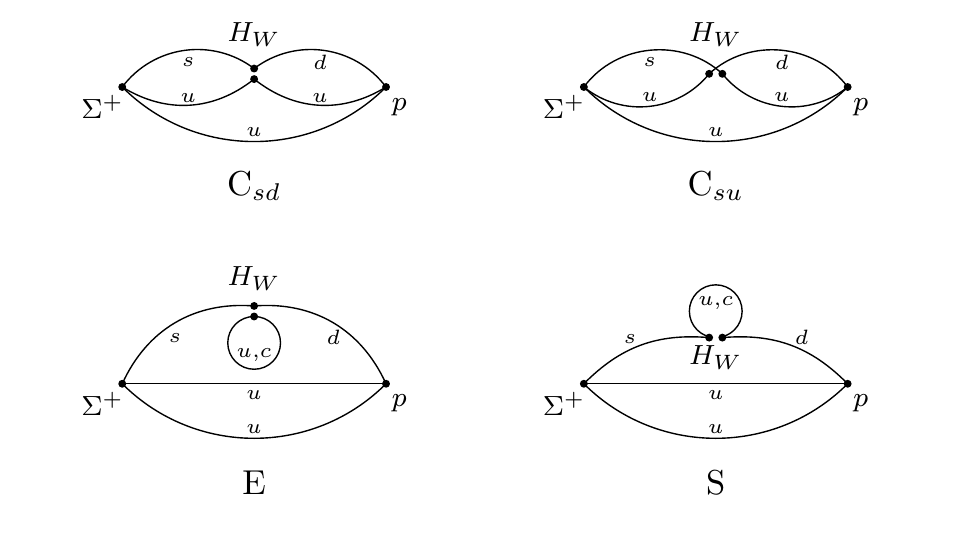}
    \caption{Four topologies of the Wick contractions of the weak Hamiltonian 3-point function. There are two connected contributions ($C_{sd}$ and $C_{su}$), and an Eye (E) and Saucer (S) diagram.}
    \label{fig:Hw_topologies}
\end{figure}

The Wick contractions for the weak Hamiltonian come in 4 different topologies shown in fig.~\ref{fig:Hw_topologies}, with the E and S type diagrams requiring a quark loop. These loops must be computed stochastically and therefore are likely to come with a large amount of statistical noise. For these proceedings, we focus only on the two connected type diagrams $C_{sd}$ and $C_{su}$, leaving the computation and analysis of the E and S diagrams to be added at a later stage.

\begin{figure}
    \centering
    \includegraphics[width=0.65\linewidth]{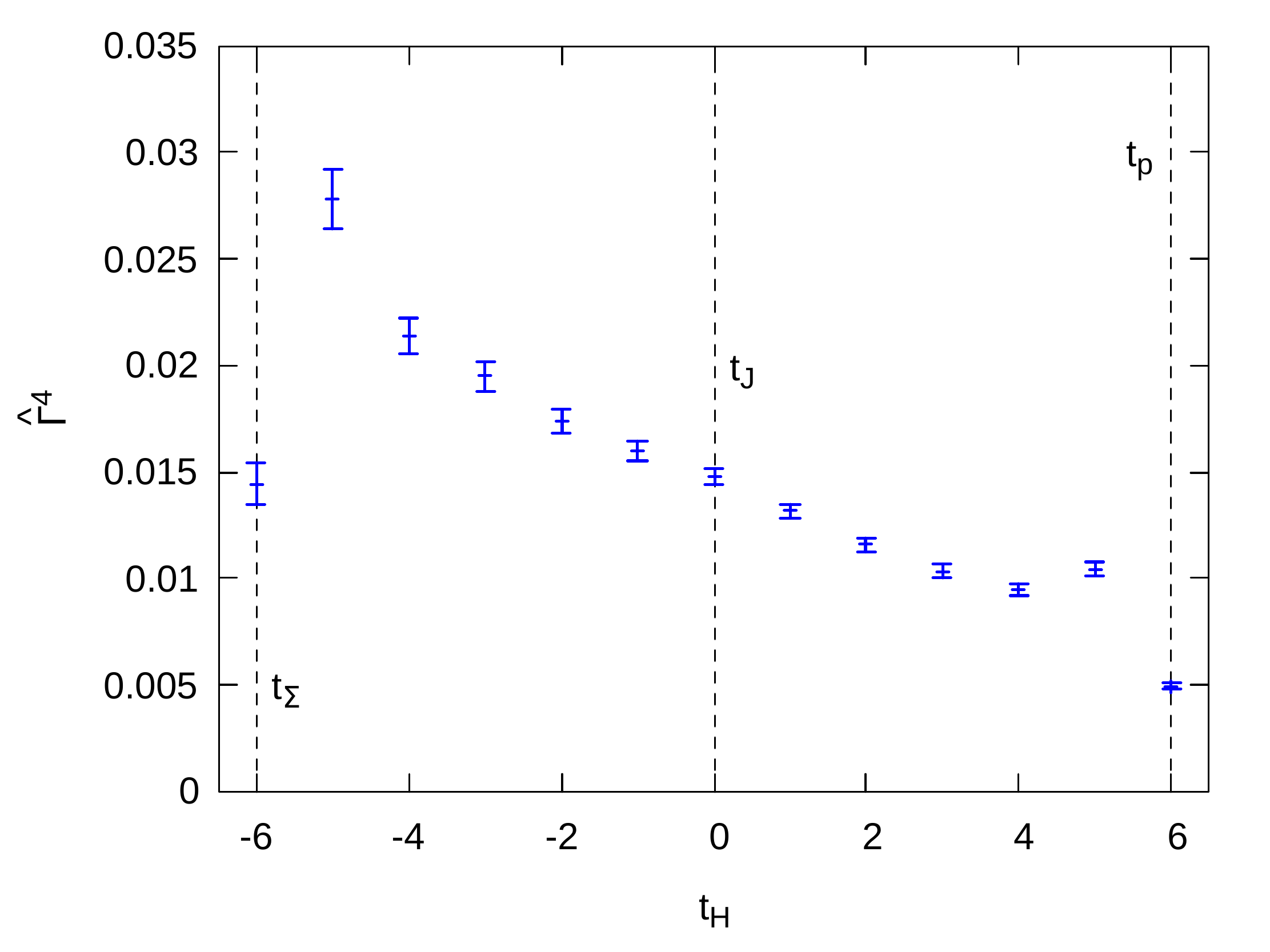}
    \caption{Trace of the amputate 4-point correlation function $\Tr[ \hat{\Gamma}^{(4)}_\mu(t_p,t_H,t_\Sigma)]$ with $t_\Sigma = -6$, $t_p = 6$ and the temporal component of the e.m. current $\mu = t$.}
    \label{fig:4pt_func}
\end{figure}

Fig.~\ref{fig:4pt_func} shows an example of the amputated 4-point correlation function in eq.~\eqref{eq:amputated_4point}, with a source-sink separation of $t_p - t_\Sigma = 12$ lattice sites, for which we are able to significantly resolve this baryonic 4-point correlation function from zero. Note that none of the intermediate state contributions have been removed. Due to this small source-sink separation, there can be large excited state contributions, as well as having a very short range of $T_a,T_b$ values to take the $T_a,T_b \to \inf$ limit. Therefore, to get some control over excited state contamination and have a comfortable range of $T_a,T_b$ values, we shall be adding increased source-sink separations to our analysis.
In addition, fig.~\ref{fig:4pt_func} only shows the temporal component of the e.m. current ($\mu = t$), however, at least one additional current direction is required to extract the form factors, as is described in ref.~\cite{Erben:2022tdu}.

\section{Conclusions and Outlook}
In conclusion, in these proceedings, we have provided an overview of the methods required to extract the hadronic amplitude of the rare Hyperon decay $\Sigma^+ \to p \ell^+ \ell^-$ from lattice QCD. This includes the removal of exponentially growing intermediate states, as well as the power-like finite volume corrections induced by these states. For full details about this work, see ref.~\cite{Erben:2022tdu}.

Also, we have presented the first preliminary data of the connected type contributions to the amputated 4-point correlation function $\hat{\Gamma}_\mu^{(4)}$ on an ensemble with unphysically heavy pions of mass $340 \,\MeV$.
An analysis of the data is in progress, and additional data is being taken for larger source-sink separations, an additional e.m. current component, and the E and S diagrams. 

Looking beyond this first exploratory calculation, we intend to move towards a physical point calculation of this decay. This comes with many challenges including the worsening of the signal-to-noise problem, decorrelation of the charm and light loop propagators in the E and S diagrams (see \cite{Boyle:2022ccj} for more information), and the requirement of an $N\pi$ scattering study as well as the $\Sigma^+ \to N \pi$ and $N \pi \to p$ transitions in the finite volume.

\acknowledgments
The authors thank the members of the RBC and UKQCD Collaborations for helpful discussions and suggestions. 
This work used the DiRAC Extreme Scaling service at the University of Edinburgh, operated by the Edinburgh Parallel Computing Centre on behalf of the STFC DiRAC HPC Facility (\href{www.dirac.ac.uk}{www.dirac.ac.uk}). This equipment was funded by BEIS capital funding via STFC capital grant ST/R00238X/1 and STFC DiRAC Operations grant ST/R001006/1. DiRAC is part of the National e-Infrastructure. 
F.E., V.G., M.T.H and A.P. are supported in part by UK STFC grant ST/P000630/1. 
F.E., V.G., R.H and A.P. also received funding from the European Research Council (ERC) under the European Union’s Horizon 2020 research and innovation programme under grant agreements No 757646 \& A.P. additionally by grant agreement 813942. Additionally, M.T.H. is supported by UKRI Future Leaders Fellowship MR/T019956/1.

\bibliographystyle{JHEP}
\bibliography{Refs}

\end{document}